\documentstyle[pra,aps,multicol,epsfig]{revtex}

\tighten

\begin{document}
\draft
\title{Collapses and revivals of exciton emission in a semiconductor\\
microcavity: detuning and phase-space filling effects}
\author{Guang-Ri Jin$^{1}$, Wu-Ming Liu$^{1,2}$}
\address{$^1$National Lab of Magnetism, Institute of Physics,
Chinese Academy of Sciences, Beijing 100080, China}
\address{$^2$Department of Physics, University of Toronto, Ontario
M5S 1A7, Canada}
\date{\today }
\maketitle

\begin{abstract}
We investigate exciton emission of quantum well embedded in a
semiconductor microcavity. The analytical expressions of the light
intensity for the cases of excitonic number state and coherent
state are presented by using secular approximation. Our results
show that the effective exciton-exciton interaction leads to the
appearance of collapse and revival of the light intensity. The
revival time is twice compared the coherent state case with that
of the number state. The dissipation of the exciton-polariton
lowers the revival amplitude but does not alter the revival time.
The influences of the detuning and the phase-space filling are
studied. We find that the effect of the higher-order
exciton-photon interaction may be removed by adjusting the
detuning.
\end{abstract}

\pacs{PACS Numbers: 42.50. Fx, 71.35.-y}


\begin{multicols}{2}

\section{Introduction}

With the development of crystal growth techniques, people now can
fabricate multi-dimensional confined nanostructure materials, such
as quantum wells, quantum lines and quantum dots. Some interesting
phenomena not observed in bulk material may take place within
these systems. The optical properties of microcavity containing
semiconductor quantum wells have been studied intensively in the
recent years \cite{Yamamoto} since the first observation of
polaritons splitting in strong-coupling regime \cite{Weisbuch}.
The concept of the exciton-polariton was originally proposed by
Hopfield \cite{Hopfield}. In an infinite bulk crystals, the
exciton is dressed by a photon with the same wave vector to form
stable polariton due to the transitional symmetry of the total
system.

It was shown that, for sufficiently small decay rates $\gamma
_{ex}$ of the excitons and $\gamma _{c}$ of the photons, the
coherent exciton-photon interaction gives rise to a periodic
exchange of energy between the exciton and the photon modes.
Therefore, the emission from the microcavity will show Rabi-like
oscillating behavior \cite{Norris,Jacobson,HCao,HWang}. By
regarding the excitons as a boson (i.e, harmonic approximation)
and neglecting exciton-exciton interaction, i.e., within the
completely linear regime, theoretical calculation of the light
intensity gave a good agreement with the observed time-domain
emission from the microcavity \cite{Jacobson,Pau}. Other effects,
such as disorder-induced inhomogeneous broaden of the excitons
\cite{Wang} and the influence of squeeze vacuum of the photons
\cite{Erenso} were also shown to have a strong influence on
coherent dynamics of the exciton-photon coupling in microcavities.

Besides the coherent interaction between the excitons and the
photons, nonlinear interaction between the excitons may play an
indispensable role on the coupled exciton-photon system. In fact,
the harmonic approximation is valid for the case that the exciton
density is much lower than the Mott density, i.e.,
$n_{ex}a_{ex}^3\leq 10^{-2}$, where $n_{ex}$ is the exciton
density and $a_{ex}$ is the two-dimensional Bohr radius. If the
exciton density becomes relative higher, the ideal bosonic model
of the excitons is no longer adequate. In this case, some residual
Coulomb interactions among excitons and the phase-space filling
effect should be taken into account \cite{Hanamura}. It is well
known that these complex nonlinear interactions lead to parametric
amplification of an incident light from the microcavity \cite
{Savvidis,Ciuti,Meissin,Saba}.

In Refs. \cite{Liu0,Liu1,Liu2}, the authors studied the effect of
the nonlinear interactions on the fluorescence spectrum of
excitons. The deviation of high density excitons from the ideal
boson model was investigated by introducing the concept of
q-deformed excitons \cite{Liu1}. With the achivements of previous
works mentioned above, it is natural to ask the effects of the
exciton-exciton interaction on the light intensity from the
semiconductor microcavity. Our previous work shows that the
nonlinear interaction will lead to the appearance of collapses and
revivals (CRs) in the light intensity \cite{CRs}. Compared the
initial coherent state case with the number state case, the
revival time is twice. However, in our obtaining the simple but
interesting relation, we ignored some important effects, such as
the effects of the quantum dissipation processes, the higher-order
exciton-photon interaction and the detuning between the exciton
and photon modes.

In this paper, we study further the coherent dynamics of the
coupled exciton-photon system in the semiconductor microcavities.
The effects not considered in our previous paper will be taken
into account. We would like to answer the following two questions:
(1) does the relation of double revival time still hold after the
consideration of these effects? (2) how can we keep the relation.
This paper is ranged as the following: in section II, we give a
general theoretical model of the interaction between a single-mode
cavity field and the exctions. We present the approximate
time-evolution operators by using the so-called {\it secular
approximation}. In section III, the time-evolution of light
intensity is calculated for the case the excitons are initially in
a number state (or coherent state) and the photons are in vacuum.
Finally, a brief summary and conclusion are presented in section
IV.

\section{Theoretical Model}

The considered system is a microcavity containing a semiconductor
quantum well embedded in a high finesse cavity. We assume that the
cavity and the quantum well are ideal, and they are in an
extremely low temperature circumstance. The quantum well interacts
with cavity field via exciton, which is an electron-hole pair
bound by the Coulomb interaction. The exciton and the photon modes
are quantized along the direction normal to the microcavity. We
will consider the lowest-order mode in this direction. The
excitons with in-plane wave vector ${\bf K}$ may only be dressed
by the photons with the same wave vector due to the transitional
invariance in the plane of the microcavity .

To further simplify the model, in this paper we will consider only one mode
of photons with wave vector ${\bf K}=0$ and frequency $\omega _{c}$ very
close to the lowest $n=1s$ exciton energy level \cite{Hanamura,Vietnam}. In
fact, at extremely low temperature, the thermal momentum of the excitons is
so small that the thermalized excitons can be neglected \cite{Liu1,Liu2}.
Combining the above considerations and neglecting the spin degrees of
freedom, one can write an effective interaction Hamiltonian for the coupled
exciton-photon system as \cite{Hanamura,Vietnam}:
\begin{eqnarray}
H &=&H_{0}+H_{NL}  \nonumber \\
&=&\omega _{c}a^{\dagger }a+\omega _{ex}b^{\dagger }b+g(a^{\dagger
}b+b^{\dagger }a)  \nonumber \\
&&+Ab^{\dagger }b^{\dagger }bb-B\left( b^{\dagger }b^{\dagger }ba+a^{\dagger
}b^{\dagger }bb\right) ,  \label{eq:H}
\end{eqnarray}%
where $b^{\dagger }(b)$ are creation (annihilation) operators of
the excitons with frequency $\omega _{ex}$, and $a^{\dagger }(a)$
are the creation (annihilation) operators of the cavity field. We
assume that both of them obey the bosonic commutation relation
$[b,b^{\dagger }]=[a,a^{\dagger }]=1$. The third term stands for
the exciton-photon interaction with coupling strength $g$, which
is larger than nonlinear interaction coefficients $A$ and $B$. The
fourth term describes the effective exciton-exciton interaction
due to Coulomb interaction. The higher-order exciton-photon
interaction, the fifth term, represents the phase-space filling
effects. For small in-plane wave vectors of the excitons
and the photons, the nonlinear interaction constants $%
2A=6Ry_{ex}a_{ex}^{2}/S $ and $B=g/\left( n_{sat}S\right) $, where
$Ry_{ex}$ is the binding energy of the excitons, $S$ the
quantization area and $ n_{sat}=7/\left( 16\pi a_{ex}^{2}\right) $
is the exciton saturation density \cite{Ciuti,Tasson}. The ratio
of the exciton-exciton interaction constant $ A $ and the
phase-space filling factor $B$ may be determined by a degenerate
four-wave mixing experiment \cite{dfwm,dfwmt}. In this paper, we
assume that these two parameters are real and positive.

The dynamical evolution of the two-mode boson system described by
Eq. (\ref {eq:H}) can not be calculated in an exact way due to the
presence of nonlinear interaction $H_{NL}$. Some approximations
will be involved in theoretical calculations. In Ref.
\cite{Liu1,Liu2}, the eigenvectors and eigenvalues of the total
Hamiltonian (\ref{eq:H}) were solved by using first-order
perturbation calculations. Recently, we restudied the dynamics of
the total system by using unperturbation calculations \cite{CRs},
in which, however, the effects of the phase-space filling terms
and the detuning between the cavity mode and the exciton eigenmode
were not included at that time. In this paper, both the terms
mentioned above and quantum dissipation processes due to the
coupling with a continuum phonon mode will be taken into
accounted.

Note that the linear part of Hamiltonian (\ref{eq:H}) may be diagonized by
introducing two polariton operators:
\begin{eqnarray}
p_{1}(t) &=&-va(t)+ub(t)\text{, }  \nonumber \\
p_{2}(t) &=&ua(t)+vb(t),  \label{eq:polaritons}
\end{eqnarray}%
where $u$ and $v$ are Hopfield coefficients for the exciton and
cavity modes, respectively. We assume that the coefficients are
real and positive. The requirement of canonical transformations of
Eq. (\ref{eq:polaritons}) yields $u^{2}+v^{2}=1$, that is
\begin{equation}
[ p_{i},p_{j}^{\dagger }] =\delta _{i,j}\text{, for }i\text{, }j=1
\text{, }2\text{.}
\end{equation}%
The inverse transformations of Eq. (\ref{eq:polaritons}) are
\begin{eqnarray}
b(t) &=&up_{1}(t)+vp_{2}(t)\text{, }  \nonumber \\
a(t) &=&up_{2}(t)-vp_{1}(t).  \label{eq:ab}
\end{eqnarray}%
Substituting Eq. (\ref{eq:ab}) into the linear parts of
Hamiltonian (\ref {eq:H}), we get
\begin{equation}
H_{0}=\sum_{j=1,2}\omega _{j}p_{j}^{\dagger }p_{j},  \label{eq:H0}
\end{equation}%
where
\begin{equation}
\omega _{j}=\frac{1}{2}\left[ \omega _{ex}+\omega _{c}+(-1)^{j}\Delta \right]
\text{, for }j=1\text{, }2\text{,}
\end{equation}%
are the lower-branch ($j=1$) and upper-branch ($j=2$) polariton energies,
respectively, and $\Delta =\omega _{2}-\omega _{1}=\sqrt{\delta ^{2}+4g^{2}}$
is the splitting energy of the two polaritons. The detuning between the
cavity mode and the exciton mode is $\delta =\omega _{c}-\omega _{ex}$. In
Eq. (\ref{eq:H0}), we have let
\begin{equation}
\delta uv=g(u^{2}-v^{2}),  \label{eq:condition}
\end{equation}%
to cancel the nondiagonal terms in $H_{0}$. This condition plus the
requirement of the canonical transformations of $p_{j}$ inspire us to define
\begin{equation}
u=\sin \theta \text{, }v=\cos \theta ,  \label{eq:xita}
\end{equation}%
and tan$2\theta =-2g/\delta $. Substituting Eq. (\ref{eq:ab}) into the
nonlinear parts of Hamiltonian (\ref{eq:H}), one can obtain the effective
polariton-polariton interaction term. Therefore, we may write the total
Hamiltonian (\ref{eq:H}) in terms of the polariton operators as
\begin{eqnarray}
H_{eff} &=&\sum_{j=1,2}\omega _{j}p_{j}^{\dagger }p_{j}+A_{11}p_{1}^{\dagger
}p_{1}^{\dagger }p_{1}p_{1}  \nonumber \\
&&+A_{22}p_{2}^{\dagger }p_{2}^{\dagger }p_{2}p_{2}+2A_{12}p_{1}^{\dagger
}p_{2}^{\dagger }p_{2}p_{1},  \label{eq:effH}
\end{eqnarray}%
where
\begin{eqnarray}
A_{11} &=&u^{3}\left( Au+2Bv\right) \text{, }A_{22}=v^{3}\left(
Av-2Bu\right) ,  \nonumber \\
A_{12}&=&A_{21}=2uv\left[Auv-B\left( u^{2}-v^{2}\right) \right] .
\end{eqnarray}%
In Hamiltonian (\ref{eq:effH}), we have neglected some terms
proportional to $p_{1}^{\dagger }p_{1}^{\dagger }p_{2}p_{2}$,
$p_{1}^{\dagger }p_{1}^{\dagger }p_{1}p_{2}$ and their Hermitian
conjugate terms, which describe scattering processes between the
two polariton branches and destroy particle-number conservation
within each polariton branch. For the case of strong-coupling with
the relative larger $g$, the energy gap $\Delta $ between the two
polariton branches becomes larger, so one may safely adopt the
so-called {\it secular approximation} \cite{Vietnam,Agawal} to
ignore the particle-number-nonconservation scattering channels. In
some Refs.\cite{Kuang1,Kuang2}, the authors calculated dynamics of
a two-component Bose-Einstein condensate system by using the so
called rotating-wave approximation (RWA). Physically, the essence
of RWA is the same with the secular approximation, which is valid
in the regime of weakly nonlinearity \cite{Agawal,CRs}, i.e.,
$A\text{, }B\ll g$.

From Hamiltonian (\ref{eq:effH}), one may find that the polariton number
operators $p_{j}^{\dag }(t)p_{j}(t)$ of each branch are the constant of
motion, i.e., $p_{j}^{\dag }(t)p_{j}(t)=p_{j}^{\dag }(0)p_{j}(0)=\text{const}
$. Thus, the total particle number operator $N=\sum_{j}p_{j}^{\dag
}p_{j}=a^{\dagger }a+b^{\dagger }b$ is also time-independent. The formal
solutions of the Heisenberg equations for the polariton operators $p_{j}(t)$
are
\begin{eqnarray}
p_{j}(t)&=&\exp\left\{-i\left[ \omega _{j}-i\gamma
_{j}/2+2\sum_{k=1}^2A_{jk}p_{k}^{\dag }p_{k}\right]
t\right\}p_{j}, \label{eq:Heisenberg}
\end{eqnarray}
where $\gamma _{1}$ ($\gamma _{2}$) is the natural linewidth of
the lower- (upper-) branch of the polariton \cite{Ciuti}. These
two parameters can be measured in the reflectivity spectrum of the
microcavity\cite{Houdre}. In Eq. (\ref{eq:Heisenberg}), the
initial time operators (say, $p_{j}(0)$) have been written in the
compact form ($p_{j}$). From now on, unless we specify otherwise,
all the compact form operators stand for the operators at $t=0$.
Although $p_{j}(t)$ and its Hermitian conjugate contain
time-independent products $p_{k}^{\dag }p_{k}$, the solutions of
some measurable quantities are not trivial. In the following of
this paper we will devote ourself to calculate the light
intensity. Some novel physical results will be presented.

\section{Collapse and revival of the exciton-polariton emission}

The oscillating emission from the microcavity had been
demonstrated \cite {Jacobson,HCao}. The theoretical calculation of
the light intensity by using the harmonic approximation gave a
good agreement with the observed time-domain emission from the
microcavity \cite{Jacobson}. In our previous paper \cite{CRs}, we
shown that the influence of the nonlinear exciton-exciton
interaction may result in collapse and revival of the light
intensity. In this section we will continue our calculations to
study the effects of detuning, phase-space filling and quantum
dissipation processing on the light intensity.

For a given initial state $|\psi (0)\rangle $ of the system, the intensity
of the light field $I_{c}(t)=\langle \psi (0)|a^{\dagger }(t)a(t)|\psi
(0)\rangle $ can be obtained as
\begin{eqnarray}
I_{c}(t) &=&u^{2}\langle p_{2}^{\dagger }p_{2}\rangle
+v^{2}\langle p_{1}^{\dagger }p_{1}\rangle -uv\left[ \langle
p_{1}^{\dagger }(t)p_{2}(t)\rangle +c.c\right] , \label{eq:I0}
\end{eqnarray}%
where $\langle ...\rangle =\langle \psi (0)|...|\psi (0)\rangle $.
The initial state of the exciton-photon system is assumed as
$|\psi (0)\rangle =\left| \phi (0)\right\rangle _{ex}\otimes
\left| 0\right\rangle _{c}$, i.e., the photons are initially in
vacuum state. From Eq. (\ref{eq:I0}), we find that only the last
term is time-dependent so we need to calculate $ \langle
p_{1}^{\dagger }(t)p_{2}(t)\rangle $. With help of Eq. (\ref
{eq:Heisenberg}), we obtain
\end{multicols}
\begin{widetext}
\begin{eqnarray}
\langle p_{1}^{\dagger }(t)p_{2}(t)\rangle
&=&e^{-2i(A_{11}-A_{22})t}e^{i(\omega _{1}-\omega _{2})t}e^{-(\gamma
_{1}+\gamma _{2})t/2}  \nonumber \\
&&\times \left\langle e^{2i(A_{11}-A_{12})p_{1}^{\dagger
}p_{1}t}p_{1}^{\dagger }p_{2}e^{2i(A_{12}-A_{22})p_{2}^{\dagger
}p_{2}t}\right\rangle  \nonumber \\
&=&-e^{2i(A_{11}-A_{22})(\langle N\rangle /2-1)t}e^{i(\omega _{1}-\omega
_{2})t}e^{-(\gamma _{1}+\gamma _{2})t/2}  \nonumber \\
&&\times \left\langle e^{-2i\theta
J_{y}}e^{2i(A_{12}-A_{11})J_{z}t}J_{-}e^{2i(A_{12}-A_{22})J_{z}t}e^{2i\theta
J_{y}}\right\rangle ,  \label{eq:p1p2}
\end{eqnarray}%
\end{widetext}
\begin{multicols}{2}
\noindent where $\langle N\rangle$ is the initial excitation
number in the microcavity and $\theta $ is defined in Eq.
(\ref{eq:xita}). In Eq. (\ref{eq:p1p2}), we have introduced the
Schwinger's angular momentums: $J_{z}=\frac{1}{2}(b^{\dagger
}b-a^{\dagger }a)$ and the ladder operator $J_{+}=\left(
J_{\_}\right) ^{\dagger }=b^{\dagger }a$, so
\begin{eqnarray}
p_{j}^{\dagger }p_{j} &=&N/2+(-1)^{j}e^{-2i\theta J_{y}}J_{z}e^{2i\theta
J_{y}}, \nonumber\\
p_{1}^{\dagger }p_{2} &=&-e^{-2i\theta J_{y}}J_{-}e^{2i\theta
J_{y}},
\end{eqnarray}%
where the total particle number operator $N$ commutes with the
introduced angular momentum operators $J_{\nu}$ ($\nu=x,y,z$). In
the derivation of the final form of Eq. (\ref{eq:p1p2}), we have
also used a relation
\[
e^{2i\theta J_{y}}\exp \left[ \lambda e^{-2i\theta
J_{y}}J_{z}e^{2i\theta J_{y}}\right] e^{-2i\theta J_{y}}=\exp
\left[ \lambda J_{z}\right] .
\]
After the introduction of the angular momentum operators, any
quantum states of the coupled exciton-photon system can be written
in terms of the angular momentum states $ |j,m\rangle
=\frac{(b^{\dagger })^{j+m}(a^{\dagger })^{j-m}}{\sqrt{
(j+m)!(j-m)!}}|0\rangle$, which is a direct product of two number
states with $j+m$ excitons in the quantum well and $j-m$ photons
in the cavity, respectively.

\subsection{Number state case}

If the excitons are initially in a number state $\left| \phi
(0)\right\rangle _{ex}=\left| N\right\rangle _{ex}$ and the
photons are initially in vacuum state, then the initial state of
the total system can be written in terms of the angular momentum
states as $|\psi (0)\rangle =|j,j\rangle $ with $j=N/2$.
Substituting this initial state into Eqs. (\ref{eq:p1p2}) and
(\ref{eq:I0}) we get
\end{multicols}
\begin{widetext}
\begin{eqnarray}
I_{c}(t) &=&\frac{N}{2}\sin ^{2}(2\theta )-\frac{N}{2}\left\{ \frac{\sin
^{2}(2\theta )}{2}e^{i\Delta t}e^{-(\gamma _{1}+\gamma
_{2})t/2}e^{2i(A_{11}-A_{12})(N-1)t}\right.   \nonumber \\
&&\left. \times \left[ \sin ^{2}\theta +\cos ^{2}\theta
e^{2i(2A_{12}-A_{11}-A_{22})t}\right] ^{N-1}+c.c.\right\} ,
\label{eq:number}
\end{eqnarray}%
\end{widetext}
\begin{multicols}{2}
\noindent where we have used the matrix elements $d_{m,m^{\prime }}^{j}(\phi
)=\langle j,m|\exp (-i\phi J_{y})|j,m^{\prime }\rangle $ and the relations
\begin{eqnarray*}
d_{j,m}^{j}(\phi )\! &=&\!(-1)^{j-m}\!\left( \matrix{2j \cr j+m }\right)
^{1/2}\!\!\left( \cos \frac{\phi }{2}\right) ^{j+m}\!\left( \sin \frac{\phi
}{2}\right) ^{j-m},  \nonumber \\
d_{j,m}^{j}(\phi ) &=&-\frac{\sin (\phi /2)}{\cos (\phi /2)}\left(
\frac{ j+m+1}{j-m}\right) ^{1/2}d_{j,m+1}^{j}(\phi ).
\end{eqnarray*}%
If we consider the resonant case $\delta =0$ (i.e., $\theta =\pi
/4$), Eq. ( \ref{eq:number}) may be reduced as
\begin{eqnarray}
I_{c}(t) &=&\frac{N}{2}\left\{ 1-\cos \left[ (2g+B(N-1))t\right] \right.
\nonumber \\
&&\left. \times \left[ \cos \left( \frac{At}{2}\right) \right]
^{N-1}e^{-(\gamma _{1}+\gamma _{2})t/2}\right\} .  \label{eq:resonant}
\end{eqnarray}%
From Eq. (\ref{eq:resonant}), we find that, besides the coherent
exciton-photon oscillating term, there are two additional terms,
i.e., a slow-varying part $\left[ \cos \left( \frac{At}{2}\right)
\right] ^{N-1}$ and an exponential decay term. The appearance of
the envelope function in Eq. (\ref{eq:resonant}) will lead to the
CRs of the light intensity. To see more clearly, we plot Eq.
(\ref{eq:resonant}) in Fig. 1 for $N=2$ and $N=11$. In the figure,
purely from the viewpoint of theoretical considerations, we take
$A(B)\sim 0.01g$ to satisfy the requirement of the
weakly-nonlinearity and $\gamma _{1}=\gamma _{2}=\gamma \sim
0.001g$. We find that the phenomena of CRs become more pronounced
with the increase of exciton number. More specially, the collapse
time depends strongly on the initial excitation number and becomes
smaller with the increase of $N$. The temporal decay of the
polaritons will result in the reduction in the revival amplitude
but does not alter to the revival time, which may be determined
only by the exciton-exciton interaction constant $A$ (see Ref.
\cite{CRs}). For the resonant case, the phase-space filling factor
$B$ may only change the energy oscillation frequency (see Eq.
(\ref{eq:resonant})). However, what we considered here is the
weakly-nonlinearity, that is $NB<<g$ , so the effect of $B$ is
very small.

It should be pointed out that at present experimental condition
the linewidth-to-Rabi frequency ratio $\gamma/g$ is about 0.1
\cite{Houdre}. For this case one may not observe the revival of
the light intensity within the lifetime of the polaritons. In
order to lower the radio one may improve the Rabi frequency. As we
know, the Rabi frequency measured in a III-V (GaAs) based
microcavity structure can be $9.4$ meV \cite{Rabi}.
\begin{figure}[tbph]
\begin{center}
\epsfxsize=8cm\epsffile{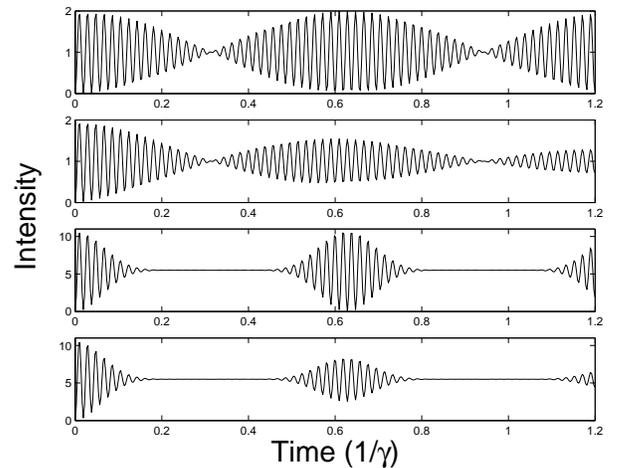} \caption{Light intensity for the
number state case as a function of time $t$. Time is in units of
$1/\gamma $ and light intensity is in an arbitrary units. The
parameters are $g=1000\gamma $, $A=0.01g$, $B=0$ and $\delta =0$.
Starting from the top: (a)$N=2$, without dissipation; (b)$N=2$,
$\gamma =1$; (c)$N=11$, without dissipation; (d)$N=11$, $\gamma
=1$.}
\end{center}
\end{figure}
The influences of the detunning and the phase-space filling are
investigated in Fig. 2. For the resonant case, the light intensity
will oscillate up-and-down around the center point $N/2$. However,
the nonresonant coupling between the excitons and the photons will
lead to the whole curve to become lower, i.e, the center line
becomes more closer to the horizontal axis. Besides, the revival
time becomes longer with the increase of the detunning. Comparing
Fig. 2 (b) with Fig. 2 (a), we find that for the case $\delta\neq
0$, the factor $B$ will increase further the revival time.
However, it is deserved to mention that if $\delta= 0$, the factor
$B$ does not give any feasible change to the revival time (see Eq.
(\ref{eq:resonant})). Our conclusion is that {\it the revival time
may be tuned by adjusting the field detuning and the phase-space
filling factor }.

\subsection{Coherent state case}
An excitonic coherent state as the initial state was used in Ref.
{\cite {Jacobson}} to stimulate the linear model solutions with
their experimental results and exhibited good agreement. So we
continue our calculations for the coherent state case. The
coherent state is characterized by $\beta = \sqrt{\langle N\rangle
}\exp (i\phi )$ with the average exciton number $ \langle N\rangle
=\left| \beta \right| ^{2}$ and the initial phase $\phi $. It
should be pointed out that the definition of excitonic coherent
state as the eigenstate of the annihilation operator of the
excitons may work well only in the low exciton density regime
\cite{jin}. This is because the expansion of a coherent state in
the number state space will involve $|N\rangle _{ex}$ with large
excitons number, which may destroy the weakly nonlinearity
condition $NA\text{, } NB\ll g$ for the fixed $g$ and $A$.
However, in the following discussions, we restrict ourself to the
weakly-excitation case with smaller average exciton numbers
$\langle N\rangle$ so one can still approximately describe the
quantum coherence natures of the exciton systems.

The initial state of the total system $|\psi (0)\rangle =\left|
\beta \right\rangle _{ex}\left| 0\right\rangle _c$ can be written
as
\begin{equation}
|\psi (0)\rangle =e^{-|\beta |^{2}/2}\sum_{j=0}^{\infty
}\frac{\beta ^{2j}}{ \sqrt{(2j)!}}\left| j,j\right\rangle.
\label{eq:coherentstate}
\end{equation}%
Substituting Eq. (\ref {eq:coherentstate}) into Eqs.
(\ref{eq:p1p2}) and (\ref{eq:I0}), we get the final result for the
light intensity at time $t$ in the coherent state representation
\end{multicols}
\begin{widetext}
\begin{eqnarray}
I_{c}(t) &=&\frac{\langle N\rangle }{2}\sin ^{2}(2\theta )-\frac{\langle
N\rangle }{2}\left\{ \frac{\sin ^{2}(2\theta )}{2}e^{i\Delta t}e^{-(\gamma
_{1}+\gamma _{2})t/2}e^{i(A_{11}-A_{22})(\langle N\rangle -1)t}e^{-\langle
N\rangle }\right.   \nonumber \\
&&\left. \times \exp \left[ \langle N\rangle
e^{-i(2A_{12}-A_{11}-A_{22})t}\left( \sin ^{2}\theta +\cos ^{2}\theta
e^{2i(2A_{12}-A_{11}-A_{22})t}\right) \right] +c.c.\right\} ,
\label{eq:coherent}
\end{eqnarray}%
\end{widetext}
\begin{multicols}{2}
\noindent For the case $\delta =0$ (i.e., $\theta =\pi /4$), Eq.
(\ref {eq:coherent}) may be reduced as
\begin{eqnarray}
I_{c}(t) &=&\frac{\langle N\rangle }{2}\left\{ 1-\cos \left[
(2g+B(\langle
N\rangle -1)) t\right]\right.   \nonumber \\
&&\left. \times e^{-2\langle N\rangle \sin ^{2}\left( At/4\right)
}e^{-(\gamma _{1}+\gamma _{2})t/2}\right\} ,  \label{eq:cohesolu}
\end{eqnarray}%
where the initial (absolute) phase of the coherent state $\phi $
does not appear in the light intensity. We find that the envelope
function for the coherent state is $e^{-2\langle N\rangle \sin
^{2}\left( At/4\right)}$, which may also lead to the CRs of the
light intensity. However, different with the number state case,
the revival time is $t=4\pi /A$, which is twice that of the number
state case (comparing Fig. 3 (a) and 3 (c) with Fig. 1 (a) and
Fig. 1(c)). Moreover, the phenomena of CRs with small average
exciton number are more pronounced for the coherent state case
than that of the number state case. In fact, the CRs can be
visible even for $\langle N\rangle <1$ due to the quantum
superposition properties of the excitonic coherent state. It is
the same with the number state case, the results of Fig. 3(b) and
Fig. 3(d) show that the decay term lowers the revival amplitude
but does not modify the revival time.

The influences of the detuning and the phase-space filling for the
coherent state case are also studied. Our results also confirm
that the detuning will change the revival time and the revival
amplitude. On the same time, the phase-space filling factor $B$
can enhance the modification of the time. However, our results
show that one may remove the influence of $B$ by adjusting the
detuning. In fact, for the resonant case, the conclusion of the
double revival time for the coherent state case compared with that
of the number state is also valid.
\begin{figure}[htbp]
\begin{center}
\epsfxsize=8cm\epsffile{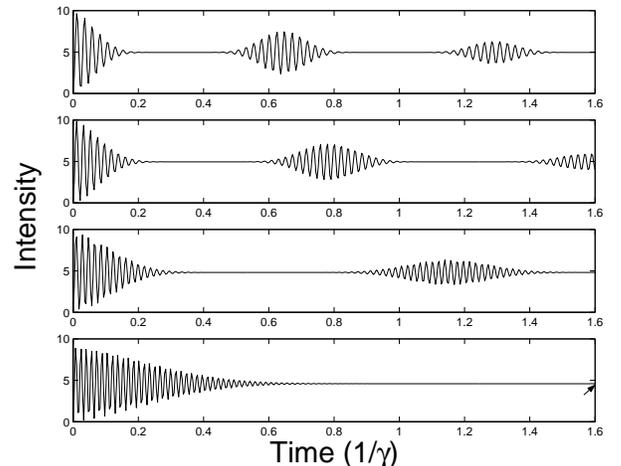}\caption{Light intensity for the
number state case as a function of time $t$, obtained from Eq.
(\ref{eq:number}). Time is in units of $1/\gamma$ and light
intensity is in an arbitrary units. Other parameters are
$g=1000\gamma$, $A=0.01g$ and $N=10$. Starting from top: (a)
$\delta=0.2g$, $B=0$; (b) $\delta=0.2 g$, $B=0.3A$; (c)
$\delta=0.4 g$, $B=0.3$; (d) $\delta=0.6 g$, $B=0.3A$ }
\end{center}
\end{figure}
To our knowledge, the CRs of Rabi oscillation in the atom-cavity
system has been studied intensely, which can be described by
Jaynes-Cummings model \cite{Eberly,Rempe}. When the single-mode
cavity field is initially in a special state (say, coherent
state), the population of the atom will exhibit CRs due to the
Rabi oscillations being modulated by different mode frequencies.
The CRs of the emission from the microcavity in the linear regime
was studied in Ref. \cite{Cao}. In their experiment, the origin of
the CRs is due to quantum beat aroused from the strong coupling of
the heavy-hole exciton and light-hole exciton state to the cavity
photon state. Within these two systems mentioned above, linear
interaction between the matter field and the light field plays a
prominent role for the appearance of the CRs. The creations of CRs
in nonlinear systems, such as the nonlinear directional coupler
\cite{CR1}, the relative phase between two superfluids or
superconductors \cite{CR2}, and the population imbalance of a
two-mode Bose-Einstein condensate
\cite{CR3,Williams,WDLi,Zhangp,Kuang1,Kuang2} have been also
studied. Here in our paper, the emission of the high-density
excitons in a quantum well embedded in a single-mode cavity is
also found to exhibit CRs due to nonlinear interaction between the
excitons.
\begin{figure}[htbp]
\begin{center}
\epsfxsize=8cm\epsffile{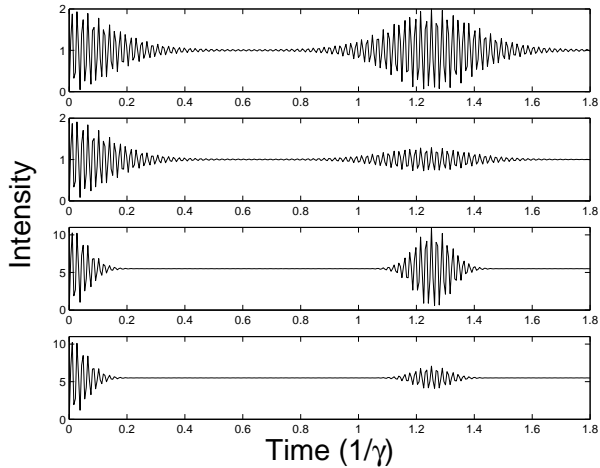}\caption{Light intensity for
coherent state case as a function of time $t$. Time is in units of
$1/\gamma$ and light intensity is in an arbitrary units. Other
parameters are $g=1000\gamma$, $A=0.01g$, $ B=0$, and $\delta=0$.
From up to down: (a) $\langle N\rangle=2$, without dissipation;
(b) $\langle N\rangle=2$, $\gamma=1$ (c) $\langle N\rangle=11$,
without dissipation; (d) $\langle N\rangle=11$, $\gamma=1$.}
\end{center}
\end{figure}
\section{conclusion and some remark}
In summary, we have studied the emission of a semiconductor
microcavity. By treating the excitons as a single-mode boson,
i.e., the harmonic approximation, the analytical expressions for
the light intensity are derived both for excitonic number state
and coherent state. The effects of the detuning of the light field
from the exciton mode, the high-order exciton-photon interaction
and the dissipation are taken into account with the help of the
secular approximation. The time evolution of the light emission is
shown to be quite different between the number state and the
coherent state cases. For the former one, the revival periods of
the oscillations are $2\pi /A$. Whereas, for the excitons in a
coherent state the revival periods are $4\pi /A$. The temporal
decay of the exciton-polariton lowers the revival amplitude but
does not modify the revival time. However, the influences of the
detuning and the phase-space filling may change both the time and
the amplitude. How can we exclude these complex modifications and
investigate only the quantum effect of excitonic states? Our
results show that, for the resonant case, the revival time is very
insensitive to the changes of the factor $B$ and the time is only
determined by the effective exciton-exciton interaction. We expect
that our theoretical study of the phenomena of collapses and
revivals would be helpful in practical experiment to measure
quantum states of excitons.

Finally, we would like to emphasize that, in our theoretical
treatment, the critical requirement of the CRs is the nonlinear
exciton-exciton interaction but not the single-mode approximation.
In fact, the treating the excitons as a single-mode harmonic
oscillator is just an idealized theoretical model. For the case of
a very strong coupling ($g\sim Ry_{ex}$), the effects of other
exciton modes (such as $\bf {K}\neq 0$) may play important role
compared with the nonlinear interactions \cite{Khurgin1,Khurgin2}.
These effects will be taken into accounted elsewhere.

This work is supported by the NSF of China under Grant Nos.
10174095 and 90103024. One of the authors (GRJ) would like to
express his sincerely thanks to Professor C. P. Sun and Dr. Y. X.
Liu.

\end{multicols}

\end{document}